\documentclass[prl, twocolumn, showpacs]{revtex4}
\usepackage{amssymb}
\usepackage{amsmath}
\usepackage{epsfig}
\usepackage{bm}
\usepackage{xcolor}

\begin{document}

\title{Negative magnetoresistance in  viscous flow \\ of two-dimensional electrons }
\author{P.\,S.~Alekseev$^*$}
\affiliation{Ioffe Physical Technical Institute,  194021  St.~Petersburg, Russia}

\begin{abstract}
At low temperatures, in very clean two-dimensional (2D) samples the
electron mean free path for collisions with static defects and
phonons becomes greater than the sample width. Under this condition,
the electron transport occurs by formation  of a viscous  flow of an
electron fluid. We study the viscous flow of 2D electrons in a
magnetic field perpendicular to the 2D layer. We calculate the
viscosity coefficients as the functions of magnetic field and
temperature. The off-diagonal viscosity coefficient determines the
dispersion of the 2D hydrodynamic waves. The decrease of the
diagonal viscosity  in magnetic field leads to negative
magnetoresistance which  is temperature- and size-dependent. Our
analysis demonstrates that the viscous mechanism is responsible for
the giant negative magnetoresistance recently observed in the
ultra-high-mobility GaAs quantum wells. We conclude that 2D
electrons in that structures in moderate  magnetic fields should be
treated as a viscous fluid.

 \pacs{72.20.-i, 
       73.63.Hs, 
       75.47.De, 
       75.47.Gk}  

\end{abstract}

\maketitle

{\bf 1.} In modern high-quality GaAs heterostructure samples with
low-temperature mobilities of 2D  electrons of the order of $10^6
-10^7$~cm$^2$/V$\cdot$s the electron mean free path for collisions
with static defects  and phonons, $\l$,  can be greater than the
sample width $w$. In this case, the transport properties depend on
the character of electron scattering at the sample edges. If the
scattering is {\em specular} and the sample has the form of a long
rectangle,  then, after several collisions with the edges, an
electron will eventually be scattered by a defect or a phonon. These
processes will determine the Drude resistivity
 $\varrho_D  = m/e^2 n \tau $, $ \tau= l / v_F, $ similar to the usual
case when $l \ll w$. Here $n$ is the electron concentration, $e$ and
$m$ are the electron charge and the effective mass, and $v_F$ is the
Fermi velocity.

If the scattering on the sample edges is {\em diffusive}, the
electron transport will be controlled by  the relation between the
mean free path for electron-electron collisions, $l_{ee} $, and the
sample width $w$. When $l_{ee}$ is greater than $w$, the scattering
at the edges dominates and the transport mean free path will be of
the order of $w$. The corresponding ``ballistic'' resistivity is
$\varrho _{ball} = m / e^2 n \tau_{ball}  $, where $ \tau_{ball}
\sim w/v_F $. In the opposite case, $l_{ee} \ll w$, the  electron
transport should resemble the Poiseuille flow in conventional
hydrodynamics with the resistance proportional to the electron shear
viscosity $\eta \sim v_F l_{ee}$. This idea was put forward (for
three-dimensional metals) by R.\,N.~Gurzhi with coauthors a long
time ago \cite{Gurzhi1, Gurzhi2, Gurzhi3}, and more recently it was
also applied to various aspects of two-dimensional electron
transport \cite{Mollenkamp1, Gurzhi4, Mollenkamp2, Mollenkamp3,
Vignale1,Sci_Rep,Andreev}. The equations describing a flow of a
viscous electron fluid in a sample have some common features with
the magnetohydrodynamic equations of charge-compensated viscous
fluids (e.g., plasma in the hydrodynamic limit)
\cite{LL8,2D_hydrodyn}.

If a sample is placed in magnetic field and the electron cyclotron
radius $R_c$ is much smaller than the sample width $w$, the
hydrodynamic regime can be realized even when $l_{ee} > w$ (but
herewith $l_{ee} \ll l $) \cite{Gurzhi2}. Indeed, an electron moving
along the trajectory similar to the circle with the radius $R_c \ll
w$ does not scatter on  the sample edges, but undergoes all other
types of scattering. The electron viscosity, like other kinetic
coefficients, becomes a tensor depending on magnetic field
\cite{LL10,Aliev}.

Another type of solid state systems with the hydrodynamic mechanism
of electron transport was studied in Ref.~\cite{Spivak}.  The
authors of that paper considered a 2D viscous electron flow
bypassing the  defects located one from another at the distances of
the order of $d \gg l_{ee} $. If the electron-electron scattering
dominates, a viscous flow in the regions between the defects is
formed and  the sample resistance is again proportional to the
viscosity $\eta $.

In this Letter we develop the hydrodynamic approach for the 2D
electron transport in magnetic field \cite{D}. We calculate the
electron viscosity tensor in a shortcut  way similar to the textbook
derivation of the Drude conductivity. The non-diagonal viscosity
$\eta_{xy}$ determines the dispersion law of the 2D hydrodynamic
waves in magnetic field. The decrease of the diagonal viscosity
$\eta_{xx}$ with magnetic field provides a mechanism for large {\em
negative} magnetoresistance of 2D electrons which is temperature-
and sample width-dependent \cite{U}. We perform detailed
calculations of magnetoresistance for the conventional Poiseuille
flow in a long rectangular GaAs sample with rough edges. We also
qualitatively demonstrate that the hydrodynamic negative
magnetoresistance arises in the  2D samples of other types, in
particular, in the samples containing large-radius defects.

The temperature-dependent giant negative magnetoresistance of 2D
electrons in high-quality  GaAs  quantum wells at low temperatures
and moderate magnetic fields, reported in several recent
publications \cite{Zudov1, Mani, Haug}, and especially the
``colossal'' negative magnetoresistance, observed in
Ref.~\cite{Zudov2}, are not understood at the present time. Several
striking features of these experiments, especially, the temperature
dependence of magnetoresistance, are in a fine agreement with the
predictions of our model. Our theory explains the existence of a
magnetoresistance  peak as well as its broadening and disappearing
with temperature
\cite{O,Another_theo_dis,Dm_Dyak_Jull,Another_theo_high_B,3D}.
Thereby we conclude that 2D electrons in the ultra-high-mobility
GaAs quantum wells in moderate magnetic fields form a viscous fluid
\cite{R}.

{\bf 2.} We recall the simple hydrodynamic approach in the extreme
case when the electron mean free path $l_{ee}$ is much less than the
2D sample width $w$, while the mean free path for scattering by
phonons and static defects is much greater than $w$. Also the sample
length $L$ is assumed to be much greater than $w$. The hydrodynamic
electron velocity $V(y)$, directed along $x$, obeys the equation:
\begin{equation}
\label{Navier_Stocks_B_0_bez_tau}
 \frac{\partial{V}}{\partial{t}}= \eta \, \frac{\partial^2V}{\partial
y^2}+\frac{e}{m}E
\:,
\end{equation}
where $\eta=v_F l_{ee}/4$ is the viscosity of the 2D degenerate
electron gas and $E$ is the electric field directed along $x$. In
the present work we neglect the compressibility and the thermal
conductivity effects.

The conventional boundary conditions require $V=0$ at $y=\pm w/2$.
This implies that the electron scattering at the sample edges is
diffusive \cite{diffusive}. In stationary case, the solution of
Eq.~(\ref{Navier_Stocks_B_0_bez_tau}) gives the parabolic velocity
profile $V(y)$. Integrating the current density $j_x(y)=enV(y)$ over
$y$ one obtains the resistivity:
\begin{equation}
\label{visc_rho}
 \rho=\frac{m}{e^2 n \tau^\star}
 \: , \;\;\;\;\; \;
  \tau^\star=\frac {w^2}{12\eta}
  \:.
\end{equation}
Here $\tau^\star$ is the ``effective'' relaxation time which, in the
hydrodynamic regime, replaces the normal momentum relaxation time
$\tau$ in the formula $\varrho_D  = m/e^2 n \tau $.

Saying precisely, by the electron-electron scattering time
$\tau_{ee}  = l_{ee}/v_F $ we have to imply the relaxation time
$\tau_{2,ee}$  of the second moment of the electron distribution
function (i.e., its harmonics $\sim e^{im\phi}$ with $m=\pm 2$,
where $\phi$ is the angle of the single electron velocity). For such
the time a calculation was done for an almost ideal Fermi gas and
the Debye model for screening of the Coulomb potential. Following
the approach of Ref.~\cite{Vignale2}, we obtained:
\begin{equation}
\label{tau_ee_Debye}
\frac{\hbar}{
 \tau_{2, ee}(T)} =A_{ee} \frac{ T^2}{ E_F }
 \:,
 \end{equation}
where $T$ is the temperature, $ E_F = m v_F^2 /2$ is the Fermi
energy, and $A_{ee}=A_{ee}(E_F)$ is a dimensionless value of the
order of 1 for $E_F$ corresponding to typical GaAs samples. However,
for that samples the electron-electron interaction energy is of the
same order of magnitude as the electron kinetic energy. Calculation
of the time $\tau_{2,ee} $ for a system of strongly interacting
electrons is very laborious, but it leads to the result
(\ref{tau_ee_Fermi_liquid}), which is quite similar to
Eq.~(\ref{tau_ee_Debye}) (see Refs.~\cite{Novikov,SM}).

\begin{figure}
\epsfxsize=200pt {\epsffile{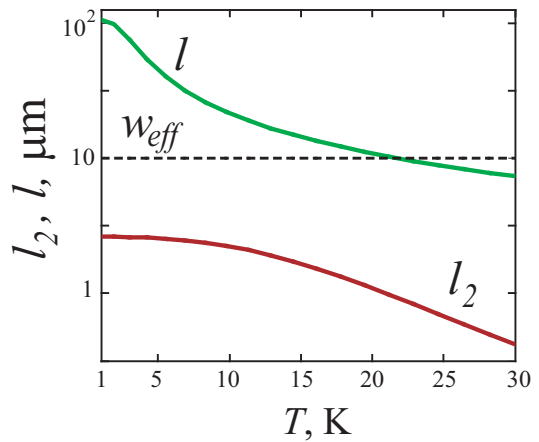}} \caption { The  mean free
paths   $l(T)$  and $ l_2(T) $ for  relaxation  of the electron
momentum and the second moment of the electron distribution
function. Calculations are performed for the sample studied in
Ref.~\cite{Zudov2}. }
\end{figure}

Thus the characteristic features of the ideal viscous electron
transport consist in (i) inverse dependence of resistivity on the
square of the sample width, $\rho\sim 1/w^2$, and (ii) inverse
dependence of resistivity on the square of temperature $T$, $\rho
\sim \eta \sim \tau_{2,ee} \sim 1/T^2 $. A hint on this very unusual
$\it{decrease}$ of resistivity with increasing temperature was
recently reported in Ref.~\cite{Zudov2} in a limited temperature
interval below 5~K.

It should be noted that generally the electron viscosity is not
necessarily related to electron-electron collisions. Any process
providing the relaxation of the second moment of the electron
distribution function (e.g., scattering on  static defects or, more
generally, on disorder) gives rise to viscosity. So the viscosity
coefficient $\eta$ is proportional to the relaxation time $\tau_2$,
for which the reciprocal value, $1/\tau_2$, contains the
contribution (\ref{tau_ee_Debye}) from the electron-electron
scattering  as well as the temperature-independent contribution from
electron scattering on disorder:
\begin{equation}
\label{tau_2}
\eta=\frac{1}{4}v_F^2 \tau_2
\:, \;\;\;\; \;\; \;
\frac{1}{\tau_2(T)}=
\frac{1}{\tau_{2,ee}(T)} + \frac{1}{\tau_{2,0}}
\:.
\end{equation}

The result given by Eq.~(\ref{visc_rho}) is modified if the momentum
relaxation time $\tau$ due to interaction with phonons and static
defects is comparable to $\tau^\star$. In this case, the usual bulk
friction term $-V/\tau$ should be added to the right-hand side of
Eq. (\ref{Navier_Stocks_B_0_bez_tau}). The modified velocity $V(y)$
profile can be easily found, and integration over $y$ gives the
following expression for the resistivity \cite{Gurzhi2,Gurzhi3}:
\begin{equation}
\label{visc_rho_tanh}
 \rho=\frac{m}{e^2 n \tau} \, \frac{1}{1 -\tanh(\xi)/\xi}
 \: , \quad  \;\; \;
 \xi=\sqrt {\frac{ 3\tau^\star }{ \tau}}
 \:.
\end{equation}
For $\tau \gg \tau^{\star}$, $\tanh\xi \approx \xi-\xi^3/3$ and the
expression for the resistivity in Eq. (\ref{visc_rho_tanh}) reduces
to Eq. (\ref{visc_rho}). In the opposite case, when $\tau \ll
\tau^{\star}$, $\tanh\xi \approx 1 \ll \xi $ and one recovers the
usual Drude resistivity $\varrho_D  = m /  e^2 n \tau$ defined by
the momentum relaxation time $\tau$.

It turns out that the following simple interpolation formula:
\begin{equation}
\label{visc_rho_s_tau_approx}
 \rho=\frac{m}{ne^2}\Big(\frac{1}{\tau}+\frac{1}{\tau^\star}\Big)
 \:,
\end{equation}
reproduces the expression (\ref{visc_rho_tanh}) for any value of
$\tau^\star/\tau$ with an accuracy better than 11\%. Thus the effect
of the electron viscosity can be regarded as a {\em parallel}
channel of electron momentum relaxation.

The values of the momentum relaxation  time $\tau_{ph}$ for
scattering of 2D electron  by acoustic phonons in the GaAs quantum
well were estimated by using the results  of Refs.~\cite{Karpus}.
According to those papers, the momentum relaxation rate is
proportional to temperature, $1/\tau_{ph} (T) =A_{ph} T $, at $T
\gtrsim  4$~K and to higher powers  of temperature at $T\lesssim
4$~K (for the structure studied in Ref.~\cite{Zudov2}).
 For the total bulk momentum relaxation rate we should
use the expression:
\begin{equation}
\label{tau_1} \frac{1}{\tau(T)}=\frac{ 1}{\tau_{ph}(T)}+
\frac{1}{\tau_{0}}
\:,
\end{equation}
where the term $1/\tau_{0}$ does not depend on temperature and is
due to electron scattering on disorder.

Fig. 1 shows the temperature dependencies of the mean free paths
$l_2 = v_F \tau_2 $ and $l = v_F \tau$ calculated  according to
Eqs.~(\ref{tau_2}), (\ref{tau_1}), (\ref{tau_ee_Fermi_liquid}), and
Refs.~\cite{Karpus} with the parameters $\tau_{2,0}$, $A_{ph}$,
$\tau_0$, $A_{ee}^{Fl}$ that will be used  further in the text to
fit the experimental data from Ref.~\cite{Zudov2}.

{\bf 3.} We now address our main point: the effects resulting from
the dependence of the electron viscosity on magnetic field.

 The
internal friction between two layers of the electron fluid moving
with different velocities is provided by the exchange of electrons
between these layers (see Fig.~2). In the absence of magnetic field
electrons from one layer penetrate into another one on a distance
which is of the order of $\l_2$ and this is what defines the
viscosity. However, in the presence of magnetic field this distance
is limited by the cyclotron radius $R_c$. Thus at strong magnetic
field the viscosity should tend to zero.

We derived the following expressions for the electron viscosity
tensor $\eta_{ij}$ \cite{SM}:
\begin{equation}
\label{eta_xx}
\eta_{xx}=  \frac {\eta   }{1+(2\omega_c \tau_2)^2}
\: ,\;\;\;
\eta_{xy}=  \frac {2\omega_c \tau_2  \eta  }{1+(2\omega_c \tau_2)^2}
\: ,
\end{equation}
where $\omega_c=eB/mc$ is the cyclotron frequency, and $\eta$ is the
viscosity at zero magnetic field introduced above. Dissipation of
energy in a viscous flow is related only to  the coefficient
$\eta_{xx}$.

The formula for the dissipative viscosity coefficient $\eta_{xx}$,
similar to the expression in Eq.~(\ref{eta_xx}), was obtained by
M.\,S.~Steinberg for a 3D metal  in Ref.~\cite{Steinberg}. The
non-diagonal viscosity $\eta_{xy}$ to our knowledge was not
considered for 2D electrons  in literature previously.

\begin{figure}
\epsfxsize=240pt {\epsffile{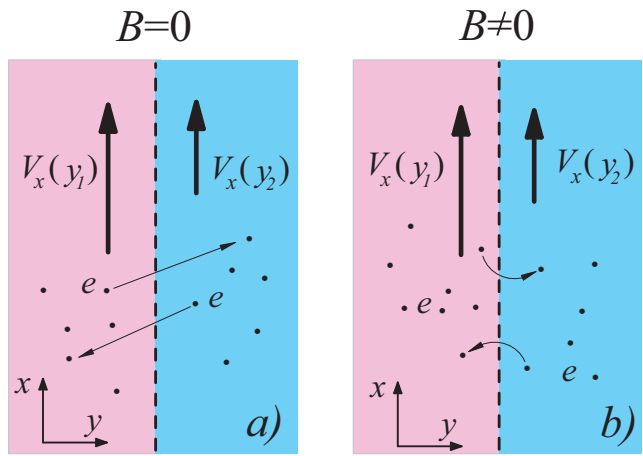}} \caption {The physical origin of
the decrease of electron viscosity in magnetic field (schematics).
Two adjacent layers of the electron fluid are moving with different
velocities $V_x(y_1)$ and $V_x(y_2)$. The viscous friction is due to
the interlayer penetration of electrons. Without magnetic field
($\it{a}$) the penetration length (defining the viscosity) is of the
order of $l_2$. In the strong magnetic field $\mathbf{B}=B
\mathbf{e}_z$  ($\it{b}$) this length is limited by the cyclotron
radius $R_c \ll l_2$. }
\end{figure}

For the hydrodynamic velocity of a 2D viscous flow in magnetic field
we derived the motion equation \cite{SM}:
\begin{equation}
\label{Navier_Stocks_B__gen___x}
\frac{\partial \mathbf{V} }{\partial t}= \eta_{xx} \Delta \mathbf{V}+ [\, (\eta_{xy}
  \Delta \mathbf{V} + \omega_c  \mathbf{V} ) \times \mathbf{e}_z \,]
  + \frac{e}{m}\mathbf{E} - \frac{\mathbf{V}}{\tau}
 \:,
\end{equation}
where $\Delta = \partial^2 / \partial x^2 + \partial^2 / \partial
y^2$. Since we neglect compressibility of the electron fluid, we
must assume that $ \mathrm{div } \mathbf{V} =0 $.

In the stationary regime and in the absence of the Hall current,
$V_y \equiv 0$, Eq. (\ref{Navier_Stocks_B__gen___x})  for a long
sample reduces to:
\begin{eqnarray}
\label{Navier_Stocks_B__flat___x}
\eta_{xx}\frac{d^2 V}{dy^2}+\frac{e}{m}E_x- \frac{V}{\tau}= 0
\:,
\\
\label{Navier_Stocks_B__flat___y}
\eta_{xy}\frac{d^2 V}{dy^2}+\omega_c V -\frac{e}{m}E_y=0
\:.
\end{eqnarray}
Here $V=V_x$,   $E_x (y)=const$ is the electric field due to the
applied voltage, and $E_y (y)$ is the Hall electric field
corresponding to the condition  $V_y \equiv 0$.

For the case of the absence of momentum relaxation in the bulk,
$1/\tau=0$, Eq. (\ref{Navier_Stocks_B__flat___x}) coincides with the
stationary version of Eq. (\ref{Navier_Stocks_B_0_bez_tau}) if one
replaces $\eta$ by $\eta_{xx}$. Thus the resistivity $\rho$ will be
given by Eq. (\ref{visc_rho}) with the additional factor
$[1+(2\omega_c \tau_2)^2]^{-1}$, describing the giant negative
magnetoresistance. For the case of a nonzero bulk momentum
relaxation rate, $1/\tau \neq 0$, the resistivity $\rho$
corresponding to Eq. (\ref{Navier_Stocks_B__flat___x}) will be
calculated by Eq. (\ref{visc_rho_tanh}), where
\begin{equation}
\label{xi_B}
\xi=\sqrt{\frac{3\tau^{\star}}{\tau} [1+(2\omega_c\tau_2)^2]}
\:,
\end{equation}
or by the approximation formula analogous to
Eq.~(\ref{visc_rho_s_tau_approx}):
\begin{equation}
\label{visc_rho_s_tau_approx__B}
\rho=\frac{m}{e^2 n } \Big( \, \frac{1}{\tau}+\frac{1}{\tau^\star}
\frac{1}{1+(2\omega_c \tau_2)^2}  \Big)
 \:.
\end{equation}
It is seen from Eq.~(\ref{visc_rho_s_tau_approx__B}) that the
decrease of $\tau_2$ and fastening of the relaxation rate $1/\tau$
with temperature leads to broadening and a shift upwards of
magnetoresistance curves (see Fig.~3). The increase of
$\tau^{\star}$ with temperature results in vanishing of negative
magnetoresistance. At low temperatures and high magnetic fields,
$\omega_c \tau_2 \gg 1$, the equations (\ref{tau_1}) and
(\ref{visc_rho_s_tau_approx__B}) yield a finite value of the
resistance, $ m/e^2 n \tau_0$, which is related only to the electron
momentum relaxation on disorder in the bulk.

The Hall  voltage can be found by integration of
Eq.~(\ref{Navier_Stocks_B__flat___y}) over $y$. The first term in
the left-hand side of Eq. (\ref{Navier_Stocks_B__flat___y}),
proportional  to the viscosity coefficient $\eta_{xy}$, is of the
order of $\omega_c (l_2/w)^2 V$ at $\omega_c \tau_2 \ll 1$ or
$\omega_c (R_c/w)^2 V$ at $\omega_c \tau_2 \gg 1$, while the second
term $\omega_c V$ is much greater.  Thus in calculation of the Hall
voltage one should take into account only the second term, and for
the Hall coefficient we obtain the usual result: $R_H=1/nec$.

The viscosity coefficient $\eta_{xy}$ is essential for
non-stationary flows. For example, it is seen from
Eq.~(\ref{Navier_Stocks_B__gen___x}) that the term proportional to
$\eta_{xy}$ gives a contribution to dispersion of the hydrodynamic
waves, while $\eta_{xx}$ is responsible for their dissipation.
Indeed, if we seek the solution of
Eq.~(\ref{Navier_Stocks_B__gen___x}) in  the wave form: $
\mathbf{V}_{\mathbf{k}}(\mathbf{r},t)
 = \mathbf{A}_{\mathbf{k}} \,
 \exp( \, - i \, \omega_k t +  i \, \mathbf{k }\cdot  \mathbf{r} )
$, assuming the absence of electric field and bulk momentum
relaxation, we easily obtain:
\begin{equation}
\label{omega_k}
\omega_k=\pm \left( \omega_c - \eta_{xy} k ^2
\right) -i \eta_{xx} k ^2 \:.
\end{equation}

\begin{figure}
\epsfxsize=260pt {\epsffile{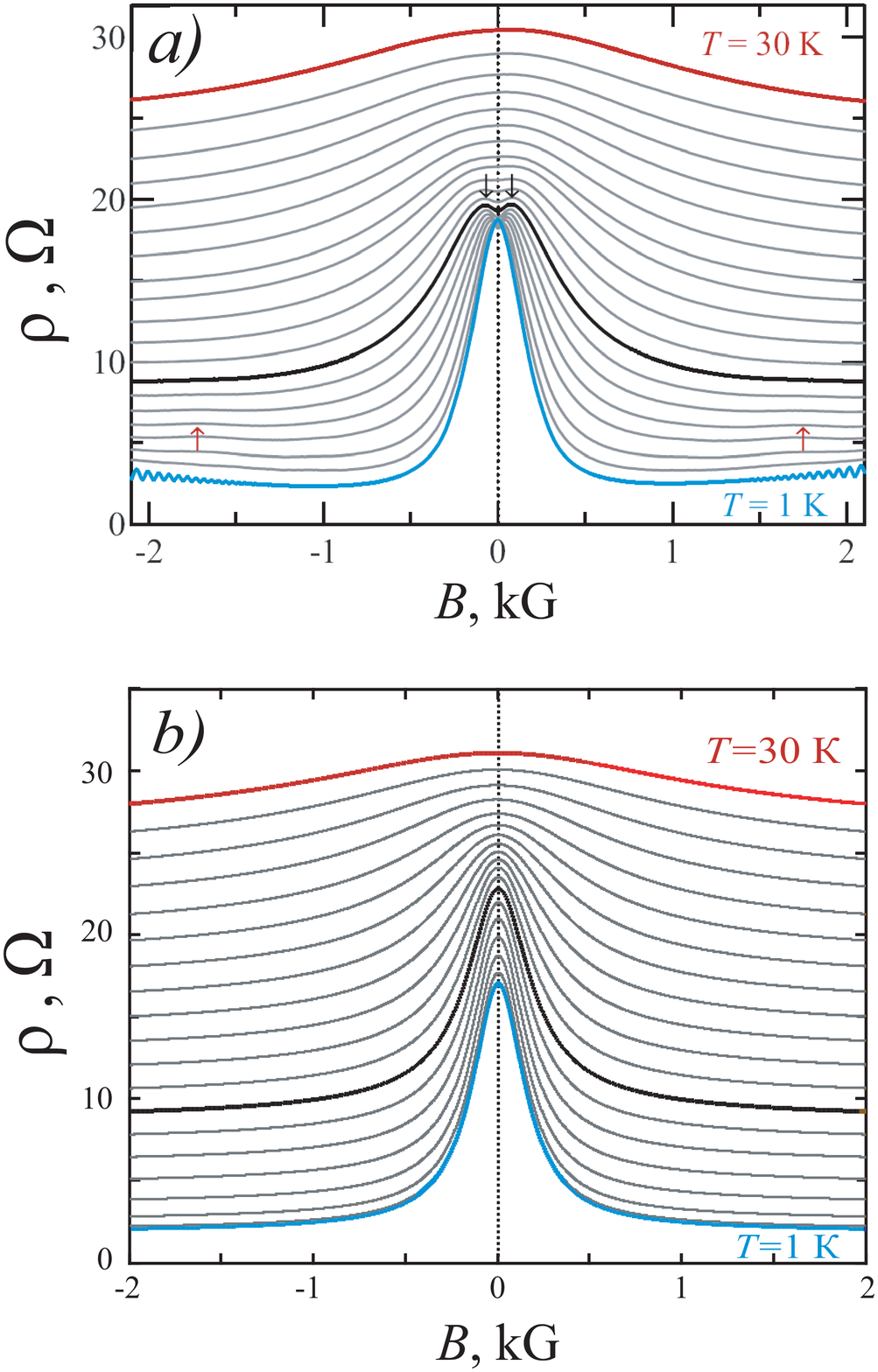}} \caption
{Temperature-dependent magnetoresistance of high-mobility 2D
electrons in the GaAs quantum well experimentally studied in
Ref.~\cite{Zudov2}. The panel (a) is taken from Ref.~\cite{Zudov2}.
The curves at the panel (b) are drawn according to
Eqs.~(\ref{visc_rho_tanh}) and (\ref{xi_B}) with the numerical
parameters presented in the main text.}
\end{figure}

{\bf 4.} We now discuss the recent experimental results on the giant
negative magnetoresistance of 2D electrons \cite{Zudov1, Mani,
Haug,Zudov2} in the light of our theory. Fig.~3(a) demonstrates the
experimental  magnetoresistance curves obtained in
Ref.~\cite{Zudov2} for an ultra-high-quality GaAs sample at
different temperatures. For the same temperatures and magnetic
fields we calculated magnetoresistance of that sample within our
theory [see Fig.~3(b)]. Herewith we used the disorder relaxation
times $\tau_{0}$, $\tau_{2,0}$ and the amplitudes $A_{ph}$,
$A_{ee}^{Fl}$ in Eqs.~(\ref{tau_1}) and (\ref{tau_ee_Fermi_liquid})
as fitting parameters.

Although by the appropriate choice of the fitting parameters we are
able to perfectly reproduce the form of the experimental curves and
their evolution with temperature, it is not possible to obtain in
such the procedure the absolute values of the sample  resistance
observed in the experiments. The only way to obtain the measured
magnitudes of resistivity within our theory is  to replace the
sample width $w$ by some {\em effective} width $w_{eff}<w$. This can
be understood in the following way. The sample contains
inhomogeneities which result in formation of the conducting channels
in the sample with the widths smaller the sample width $w$.

Indeed, in the samples where the giant negative magnetoresistance
effect was observed there often exist large-radius oval defects
arising in the process of growth of the heterostructures
\cite{Haug_Private,Bockhorn_Gornyi_Haug}. The distance $d$ between
the defects varies in the range 20-100$\mu$m, while their radii are
of the order of 20$\mu$m \cite{Haug_Private}.

In vicinities of the defects the hydrodynamic velocity
$\mathbf{V}(\mathbf{r})$ cannot has a component in the direction
perpendicular to the defect edge. A slowdown of the flow occurs due
to the viscous transfer of the  $x$ component of the electron
momentum in the $y$ direction from the regions  {\em between} the
defects to the regions which are immediately {\em in front of } the
defects (in the $x$ direction). So the large-radius defects lead to
momentum relaxation by the mechanism, analogous to the diffusive
scattering on rough sample boundaries, as well as to formation of
the conducting channels with the widths smaller than $w$. At the
scales of the order of $d$ the picture of fluid motion is rather
similar to the Poiseuille flow in a rectangular sample with the
width $w_{eff} \sim d$.  The details of the velocity field
$\mathbf{V}(\mathbf{r})$ are very complicated, but the
relationship for the averaged resistance:
\begin{equation}
\label{rho_proportional_eta_xx}
 \varrho \sim \frac{1}{\tau} + \frac{
\eta_{xx} }{d^2}
 \:,
\end{equation}
analogous to  Eq.~(\ref{visc_rho_s_tau_approx__B}), will sustain
(see Ref.~\cite{SM} for a qualitative derivation  of
Eq.~(\ref{rho_proportional_eta_xx}) and  Ref.~\cite{P} for its
rigorous derivation and analysis).

In Fig.~3(b) we drew magnetoresistance calculated with the following
fitting parameters: $\tau_{0}=4.5 \cdot 10^{-10}$~s, $\tau_{2,0}=1.1
\cdot 10^{-11}$~s, $A_{ph}=10^9 $~s$^{-1}\cdot$K$^{-1}$,
$A_{ee}^{Fl}=1.3 \cdot 10^9$~s$^{-1}\cdot$K$^{-2}$, and
$w_{eff}=$10~$\mu$m. Herewith the condition of applicability of the
hydrodynamic approach, $l_2 \ll w_{eff}$, is fulfilled at all the
temperatures (see Fig.~1). The used values of  $A_{ph}$ and
$A_{ee}^{Fl}$ are in agreement by the order of magnitude with the
result of our estimations of the parameters $A_{ph}$ and $A_{ee}$
for the quantum well studied in Ref.~\cite{Zudov2}.

{\bf 5.} In conclusion, a  hydrodynamic   mechanism for 2D electron
transport in magnetic field  has been studied. We have demonstrated
that this mechanism is responsible for the giant negative
magnetoresistance, recently observed in the ultra-high-mobility 2D
electrons in GaAs/AlGaAs heterostructures.

The author wishes to thank Prof.~M.I.~Dyakonov, under whose guidance
this research  was undertaken, for his advice and support during the
course of the work, for his participating in writing the text of the
paper. The author also thanks B.\,A.~Aronzon, A.\,P.~Dmitriev,
I.\,V.~Gornyi, V.\,Yu.~Kachorovskii, G.\,M.~Minkov, A.\,D.~Mirlin,
and D.\,G.~Polyakov for numerous fruitful discussions. The work was
supported by the Russian Fund for Basic Research (Contracts No.
16-02-01166-a, 15-02-04496-a, 14-02-00198-a),  by the Russian
Ministry of Education and Science (Contract No. 14.Z50.31.0021,
Leading scientist: M.~Bayer), by the  Russian Federation President
Grants (Contracts No. NSh-1085.2014.2 and MK-8826.2016.2), and by
the Dynasty Foundation.

{\small $\;\; ^*$Corresponding author: pavel.alekseev@mail.ioffe.ru}

\newpage

\section{Supplemental Material}

We discuss  the  temperature dependencies of the 2D electron
relaxation times in the presence of the strong electron-electron
interaction and the interaction with disorder. We present  a
Drude-like derivation of the hydrodynamic equations for a 2D
electron fluid in magnetic field as well as a qualitative
consideration of a 2D viscous flow in a sample containing
large-radius defects.

\subsection{ 1. 2D electron relaxation times   }

For the case of strongly interacting electrons, consideration of
viscous transport should be based on the Fermi-liquid theory. The
time $\tau_{2,ee}$ should be treated  as the relaxation time of the
second moment of the quasiparticle distribution function due to
quasiparticle-quasiparticle  collisions. It should be  calculated
from the quantum kinetic equation for the excitations in the 2D
electron Fermi-liquid. In Ref.~\cite{Novikov} it was shown that the
dependence $1/\tau_{2,ee}(T)$, apart from the $T^2$ factor related
to the Fermi statistics of the quasiparticles, additionally contains
the logarithmic factor (related to renormalization of scattering
probability for the pairs of quasiparticles with approximately zero
total momentum):
\begin{equation}
\label{tau_ee_Fermi_liquid}
 \frac{1}{\tau_{2, ee}(T)}=A_{ee}^{Fl} \,
  \frac{T^2}{ \left[\ln \left( E_F/T\right) \right] ^2  }
  \:.
\end{equation}
The coefficient $A_{ee}^{Fl}$ is expressed via the Landau
interaction parameters, which depend on relative magnitudes of the
electron-electron interaction energy $E_C$ and the Fermi energy
$E_F$. Unfortunately, the Landau parameters (and, thus, the value of
$\tau_{2, ee}$) cannot be found in a closed form as functions of the
interaction parameter $r_s = E_C/E_F$.

It should be noted that, even for weakly interacting  degenerate 2D
electrons, the dependence of the quantum lifetime $\tau_{0,ee}$ due
to the electron-electron scattering contains the logarithmic factor
\cite{Vignale2}:
\begin{equation}
\frac {\hbar}{\tau_{0,ee}(T)} = A_{ee}^0 \, \frac{ T^2 \, \ln(2
E_F/T) }{E_F} \:,
\end{equation}
where $A_{ee}^0$ is the  numeric constant of the order of 1. Such
the behavior of  $\tau_{0,ee}(T)$ is related to kinematics of
electron-electron collisions in the 2D case.

Disorder in a structure containing 2D electrons  can modify the
physical nature and the probability of electron-electron collisions.
In particular, when a collision of two electrons in the presence of
a disorder potential occurs, the total momentum of colliding
electrons is  not conserved. Our estimations show that the
corresponding {\em disorder-assisted} contribution to  the
electron-electron relaxation rate of the second moment of the
distribution function is quadratic by  the temperature:
\begin{equation}
\frac{1}{\tau^{da}_{2,ee} (T) }= A_{ee}^{da} \, T^2 \:.
\end{equation}
The value of the coefficient $A_{ee}^{da}$ depends on the strength
and the type of disorder.

\subsection{ 2. Drude-like derivation of the hydrodynamic equations in magnetic field}

Here we present a simple derivation of  the formulas (8) for the
electron viscosity tensor $\eta_{ij}$ in magnetic field and the
motion equation (9) for the velocity field $\mathbf{V} (\mathbf{r} ,
t) $ .

The standard method, which takes many pages of laborious
calculations,  is based on the kinetic equation for the electron
distribution function in the presence of the external fields
$\mathbf{E}$ and $\mathbf{B}$ and a space-dependent hydrodynamic
velocity $\mathbf{V} (\mathbf{r},t) $ (see, for example, Refs.
\cite{LL10,Steinberg, Aliev}). We will use a simplifying shortcut
similar to the Drude approach to the electron transport where the
momentum relaxation time $\tau$ is considered as a given parameter.
In our case this main given parameter will be the time $\tau_2$ of
relaxation of the second moment of the distribution function due to
electron-electron and disorder scattering. The only (but important)
advantage of the kinetic equation approach compared to the
Drude-like approach is that the relevant relaxation times as well as
the numerical coefficients in $\eta_{ij}$  are calculated on the
way.

The viscosity terms in the hydrodynamic equation can generally be
expressed through the viscous stress tensor (per one particle) $\Pi
_{ij} = m\langle v_i v_j \rangle$, where $\mathbf{v} =(v_x,\,v_y)$
is the 2D velocity of a single electron and the angular brackets
stand for averaging over the electron velocity distribution at a
given point $\boldsymbol{\mathit{r}} = (x,\,y)$. The motion equation
for the hydrodynamic velocity $\mathbf{V} =\langle \mathbf{v}
\rangle$ in the absence of magnetic field is:
\begin{equation}
\label{dmV__dt} m\frac{\partial V_i}{\partial t}=-\frac{\partial\Pi
_{ij}}{\partial x_j} -\frac{mV_i}{\tau}  + eE_i \:.
\end{equation}
Here and below,   summation over repeating indices is assumed. At a
time scale much greater than the relaxation time $\tau_2$ the
expression for $\Pi _{ij}$ is given by \cite{LL}:
\begin{equation}
\label{Pi_ij_0} \Pi_{ij}=\Pi_{ij}^{(0)} = -m\eta V_{ij} \:,
\quad\quad V_{ij} = \frac{\partial V_i}{\partial x_j}+
\frac{\partial V_j}{\partial x_i} \:.
\end{equation}
Using Eqs.~(\ref{dmV__dt}) and (\ref{Pi_ij_0}) and taking into
account that the electron fluid is considered as incompressible in
the present study ($ \mathrm{div} \mathbf{V}=0$), one obtains the
basic Eq.~(1) with the additional term $-V/\tau$. The value given by
Eq. (\ref{Pi_ij_0}) is attained during the time $\tau_2$, as
described by the Drude-like equation:
\begin{equation}
\label{Pi_ij_relaxation} \frac{\partial\Pi_{ij}}{\partial t}
 = -\frac{1}{\tau_2}\big(\Pi_{ij}-\Pi_{ij}^{(0)}\big)
\:.
\end{equation}

In the presence of magnetic field, additional terms will appear in
the equations for $\partial V_i / \partial t $ and $\partial
\Pi_{ij} / \partial t $, since now the quantities  $\langle v_i
\rangle$  and $\langle v_i v_j \rangle$ will change in time not only
because of collisions and the electric field, but also due to the
magnetic part of the Lorenz force: $ (\partial v_i/\partial t)_{mag}
= \omega_c\epsilon_{zik}v_k$. Here $\epsilon_{lik}$ is the unit
antisymmetric tensor and $z$ is the direction of the magnetic field
$\mathbf{B}$, which is perpendicular to the 2D electron layer.
 Thus for the additional terms in the motion equations  one obtains:
\begin{equation}
\label{Pi_ij_Newton}
\begin{array}{l}
\displaystyle \Big(\frac{\partial \langle v_i  \rangle}{\partial t}
\Big)_{mag} = \omega_c \epsilon_{zik} \langle v_k  \rangle \:,
\\
\\
\displaystyle \Big( \frac{\partial \langle v_i v_j \rangle}{\partial
t}\Big)_{mag} = \omega_c  \big( \epsilon_{zik}\langle v_k v_j
\rangle +\epsilon_{zjk}\langle v_i v_k\rangle\big) \:.
\end{array}
\end{equation}
The terms (\ref{Pi_ij_Newton}) should be added to the right-hand
side of Eqs.~(\ref{dmV__dt}) and (\ref{Pi_ij_relaxation}).

Considering the steady-state solution, we get from
Eq.~(\ref{Pi_ij_relaxation}) and (\ref{Pi_ij_Newton}) the following
relation:
\begin{equation}
\label{Pi_ij_stationary_whole}
\Pi_{ij}-\omega_c\tau_2\big(\epsilon_{zik}\Pi_{kj}+\epsilon_{zjk}\Pi_{ik}\big)=\Pi_{ij}^{(0)},
\end{equation}
allowing to find the components of the tensor $\Pi_{ij}$:
\begin{equation}
\begin{array}{c}
\label{Pi_xx_via_V_ij}
 \displaystyle
\Pi_{xx}=-\Pi_{yy}=\frac{1}{1+\beta^2}\;\Pi_{xx}^{(0)}+\frac{\beta}{1+\beta^2}\;\Pi_{xy}^{(0)}
\:,
\\
\\
 \displaystyle
\Pi_{xy}=-\Pi_{yx}=\frac{1}{1+\beta^2}\;\Pi_{xy}^{(0)}-\frac{\beta}{1+\beta^2}\;\Pi_{xx}^{(0)}
\:,
\end{array}
\end{equation}
where $\beta=2\omega_c\tau_2$. Here we used the relationship
$\Pi_{ii}=0$, which follows from $\mathrm{div } \mathbf{V} =0$ and
Eq.~(\ref{Pi_ij_stationary_whole}). The components of the viscosity
tensor are defined as the coefficients in the linear relationship
between the tensors $\Pi_{ij}$ and $V_{ij}$ (see Ref.~\cite{LL10}):
\begin{equation}
\label{def_tensor_eta} \Pi_{ij} = m \, ( \eta_{xx} V_{ij} +
\eta_{xy} \epsilon_{z ik} V_{kj} ) \:,
\end{equation}
which follows from Eqs.~(\ref{Pi_ij_0}) and (\ref{Pi_xx_via_V_ij}).

With the help of Eqs.~(\ref{dmV__dt}), (\ref{Pi_ij_Newton}),
(\ref{Pi_xx_via_V_ij}), (\ref{def_tensor_eta}), and the condition
$\mathrm{div } \mathbf{V} =0$ we arrive to Eqs. (8) and (9) that
were used in the main text.

We have checked that the conventional method based on the classical
kinetic equation gives exactly the same results as this simple
derivation.

We assumed in the main text that the zero-temperature relaxation
times $ \tau_0 $ and $ \tau_{2,0} $, related to electron scattering
on disorder, do not depend on magnetic field. For example, this is
the true for electron scattering on the isolated static defects if
the electron collisions with defects are  described by classical
mechanics and the cyclotron radius is much greater than the distance
between neighbor defects \cite{Dm_Dyak_Jull}. However, for many
types of disorder and intervals of magnetic field, 2D electron
kinetics  can be considered only quantum-mechanically. In this case,
for example,  the bulk conductivity tensor even in non-quantizing
magnetic fields, $ \hbar \omega_c \ll T $, is not given by  the
Drude formulas
\cite{Dmitriev_Mirlin_Polyakov_Zudov,Vavilov_Aleiner}. The motion
equations, analogous to  Eqs.~(\ref{dmV__dt}) and
(\ref{Pi_ij_relaxation}),  should be derived from the quantum
kinetic equation. In the resulting quantum hydrodynamic equation,
the magnetic field dependencies of the kinetic coefficients
(analogous to $ \eta_{xx} $, $ \eta_{xy} $, and $ 1/\tau_{0} $) will
be determined by the type of the disorder and temperature.

\subsection{ 3. 2D viscous  flow in a sample with large-radius defects }

The hydrodynamic mechanism of electron transport can be realized in
the 2D high-mobility  samples containing isolated widely spaced
defects \cite{Spivak,Andreev}. A viscous  flow of electron fluid
bypassing  the  defects will be formed if the mean distance between
the defects, $d$, is enough large: $d \gg l_{2}$. The 2D electron
fluid is confined by the sample edges as well as by the edges of the
defects (thus the boundary  of the conducting subregion of the
sample has a complicated geometry).

For the case of the absence of magnetic field such the model was
studied in Refs.~\cite{Spivak,Andreev}. It was obtained that
resistivity of the  system is proportional to the viscosity:
\begin{equation}
\label{rho_proportional_eta_Spivak_0} \varrho \sim \eta \:,
\end{equation}
when  the radius   of the defects, $r_0$, is large: $r_0 \gg l_{2}$;
or to the viscosity  with  a logarithmic factor:
\begin{equation}
\label{rho_proportional_eta_Spivak} \varrho \sim \frac{\eta} {
\displaystyle \ln\left( d / \, l_{2}   \right)} \:,  \;\;\;\;\;\;
\; l_2 \sim  \eta \:,
\end{equation}
when the radius   of the defects is small: $r_0 \ll l_{2}$. Up to
the logarithmic factor, which cannot be very large for real samples,
the results  (\ref{rho_proportional_eta_Spivak_0})  and
(\ref{rho_proportional_eta_Spivak}) are similar to the formula (2),
obtained for the Poiseuille flow in a long rectangular sample.

The exact consideration of a 2D flow in the sample with the
large-radius defects in the presence of magnetic field must be done
on base of Eq.~(9) with the corresponding boundary conditions on
$\mathbf{V}(\mathbf{r})$ and $\mathbf{E}(\mathbf{r})$ at the sample
edges and at the defect edges. We believe that  the relationship
(15), leading to the giant negative magnetoresistance, will be
obtained in a wide range of the parameters $d$, $r_0$, $l_2$. Below
we give a qualitative consideration of such the flow for simplest
case  when the radius of defects is of the same order of magnitude
as the mean distance between them:  $r_0 \sim d \gg l_{2} $.

We study the electron flow in a long rectangular sample  with the
width $w\gg d$ and the length $L \gg w$. Following to
Ref.~\cite{Dykhne}, we introduce the mean values of velocity
$\mathbf{\overline{V}}$ and electric field $\mathbf{\overline{E}}$,
which are the results of averaging of the values
$\mathbf{V}(\mathbf{r})$ and $\mathbf{E}(\mathbf{r})$ over  a volume
with the size much greater than the character distance between
defects, $d$. The result of  such averaging of the value $\Delta
\mathbf{V}$ can be estimated as follows:
  \begin{equation}
\overline{\Delta \mathbf{V}}  \sim   -
\frac{\mathbf{\overline{V}}}{d^2}
  \:.
  \end{equation}
Here we make use of spatial homogeneity of the system on the scales
much  greater that $d$ and an approximately oscillating character of
the dependence  $\mathbf{V}(\mathbf{r})$ with the characteristic
period of the order of   $d \sim r_0$.

As $L\gg w$, the component  of the averaged electric field along the
sample $\overline{E}_x$ is equal to  the electric field from the
applied voltage and the averaged velocity $\mathbf{\overline{V}}$
has only the component  $\overline{V}_x$ along the sample direction.
The $x$ component of the Navier-Stokes equation  (9) takes the form:
\begin{equation}
\label{V_bar} -\eta_{xx}\frac{\overline{V}_x}{d^2}
 +\frac{e}{m}\overline{E}_x- \frac{\overline{V}_x}{\tau}= 0
\:,
\end{equation}
which immediately leads to the magnetoresistance (13) and (15). Let
us remind  here that Eq.~(13) is just the result of mathematical
interpolation of the exact formula (5), obtained for the
conventional Poiseuille flow in the flat sample. So Eq.~(13) is {\em
more general} than the assumptions about the geometry of the flow
used for its derivation.

In formation of a viscous flow in a sample with the large-radius
defects, the character of electron scattering (diffusive or
specular)  on the sample edges and the defect edges is not
essential. The nature of momentum relaxation in such the system is
following. In vicinities of the defects the hydrodynamic velocity
cannot has a component perpendicular to the defect surface. A
slowdown of the flow occurs due to the viscous transfer of the $x$
component of the momentum in the $y$ direction between the regions
 which are immediately {\em in front of} the defects (in the $x$ direction)
 and the regions  {\em between} the defects.

Comparing Eqs. (2) and (\ref{V_bar}), we conclude that the time
$\tau^{\star} = d^2 / \eta_{xx}$ can be interpreted  as the {\em
non-local} momentum relaxation time due to the viscosity effect. In
other words,  the time $\tau^{\star}$  describes the momentum
relaxation in the systems which are characterized by the two
features: (i) spatial inhomogeneity of the electron momentum
relaxation  rate [related to bypassing the large-radius defects or
to diffusive scattering  on the rough sample edges]; (ii) viscous
transfer of the mean electron momentum
 due to inhomogeneity  of the velocity field $\mathbf{V}(\mathbf{r})$.

Besides that, typical GaAs 2D samples often have a complicated
geometry of their edges (for example, see the sample image in
Ref.~\cite{Mani}). Irregularities of the sample edges may lead to a
slowdown of a viscous flow by the way similar as it was described
above for the viscous flow in a sample with the large-radius
defects.

We believe that the relative contributions from the processes of
diffusive scattering on the sample edges and from bypassing the
large-radius defects in formation of a viscous flow significantly
vary from sample to sample, resulting in  a variety of observing
features of the giant negative  magnetoresistance effect (see
Refs.~\cite{Zudov1,Mani,Haug,Zudov2}).


\begin{thebibliography}{}
\bibitem{U} For the first time, the hydrodynamic mechanism for negative magnetoresistance of 2D electrons
was discussed in Ref.~\cite{Gurzhi4}. However, the authors of
Ref.~\cite{Gurzhi4} believed that such the mechanism is unlikely for
existing  GaAs quantum wells.
\bibitem{O} Other existing mechanisms for strong negative
magnetoresistance (see
Refs.~\cite{Another_theo_dis,Dm_Dyak_Jull,Another_theo_high_B,3D})
cannot explain the experiments \cite{Zudov1, Mani, Haug,Zudov2}.
Indeed, the mechanisms proposed in
Refs.~\cite{Another_theo_dis,Dm_Dyak_Jull} are based on the memory
effects in scattering of 2D electrons on disorder, and therefore the
resulting magnetoresistance is independent on temperature in
contrast to the  experimental data. The
work~\cite{Another_theo_high_B}  deals with the
quasi-two-dimensional systems in quantizing magnetic field, which
are significantly different from the GaAs systems studied in
Refs.~\cite{Zudov1, Mani, Haug,Zudov2}.
\bibitem{diffusive}
Electron scattering on the edges can be treated as diffusive when
the sample edges are rough   or when the conducting subregion of the
sample has a complicated geometry; see some discussion of this topic
in Section 4 and in Refs.~\cite{SM,P}.
\end{thebibliography}
\end{document}